\documentclass[journal,a4paper]{IEEEtran}

\usepackage[pdftex]{graphicx}
\usepackage{amsmath,bm}
\usepackage{amssymb}
\usepackage{color,soul}

\usepackage{cite}

\graphicspath{{figures/}}

\begin{document}
\title{Reduction of Joule Losses in Memristive Switching Using Optimal Control}

\author{Valeriy~A.~Slipko and Yuriy~V.~Pershin,~\IEEEmembership{Senior~Member,~IEEE}
\thanks{V.~A.~Slipko is with the Institute of Physics, Opole University, Opole 45-052, Poland (e-mail: \mbox{vslipko@uni.opole.pl})}
\thanks{Y.~V.~Pershin is with the Department of Physics and Astronomy, University of South Carolina, Columbia, SC 29208 USA (e-mail: \mbox{pershin@physics.sc.edu})}
\thanks{Manuscript received April 22, 2024; revised December 6, 2024}}

\maketitle

\begin{abstract}
 This theoretical study investigates strategies for minimizing Joule losses in resistive random access memory (ReRAM) cells, which are also referred to as memristive devices. Typically, the structure of ReRAM cells involves a nanoscale layer of resistance-switching material sandwiched between two metal electrodes. The basic question that we ask is what is the optimal driving protocol to switch a memristive device from one state to another.  In the case of ideal memristors, in the most basic scenario, the optimal protocol is determined by solving a variational problem without constraints with the help of the Euler-Lagrange equation. In the case of memristive systems, for the same situation, the optimal protocol is found using the method of Lagrange multipliers. We demonstrate the advantages of our approaches through specific examples and compare our results with those of switching with constant voltage or current. Our findings suggest that voltage or current control can be used to reduce Joule losses in emerging memory devices.
\end{abstract}



\section{Introduction} \label{sec:Intro}

The problem of memristive switching optimization has a few facets. Minimizing Joule heat, switching time, or a combination of the two are possibilities. Furthermore, memristive devices can be described using ideal models~\cite{chua71a}, memristive models~\cite{chua76a,pershin09b,kvatinsky2012}, or probabilistic models~\cite{dowling2020probabilistic,DowlingSPICE,Ntinas21a,Slipko23jumps}. Lastly, the optimization problem can be formulated with application-specific constraints and/or physics-based constraints.
The highest voltage that can be used in the circuit or the compliance current to prevent the device from being damaged are the constraint examples.

To proceed, we shall first introduce the memristive devices~\cite{chua76a} and their certain subset known as ideal memristors~\cite{chua71a}.
The voltage-controlled memristive devices are defined by the set of equations~\cite{chua76a}
\begin{eqnarray}
I(t)&=&R^{-1}_M\left(\boldsymbol{x},V \right)V(t), \label{eq1}\\
\dot{\boldsymbol{x}}&=&\boldsymbol{f}\left(\boldsymbol{x},V\right), \label{eq2}
\end{eqnarray}
where $V$ and $I$ are the voltage across and current through the device, respectively, $R_M\left( \boldsymbol{x}, V\right)$ is the memristance (memory resistance), $\boldsymbol{x}$ is a vector of $n$ internal state variables, and $\boldsymbol{f}\left(\boldsymbol{x}, V\right)$ is a vector state evolution function. Current-controlled memristive devices are defined similarly~\cite{chua76a}. In the above set, Eq.~(\ref{eq1}) is the generalized Ohm's law, while Eq.~(\ref{eq2}) is the state equation. The latter defines the evolution of the internal state variable or variables.

In ideal memristors~\cite{chua71a}, the state evolution function is often proportional to the current. Consequently, the internal state variable $x$ is proportional to the charge $q$. In addition, it is common to set the proportionality coefficient to one. In this case, the internal state variable is simply the charge flown through the device from an initial moment of time. In what follows, the response of such \textit{ideal} devices is described in terms of a generalized Ohm's law
\begin{equation}
    V=R_M\left(q \right)I(t), \label{eq:ideal}
\end{equation}
where the memristance, $R_M$, is a function of charge.
Although this model is quite abstract (physical devices behave substantially differently from the ideal ones~\cite{pershin18a,kim2020experimental,di2023memristors}), its straightforward structure is advantageous for analytical calculations.


In this paper, we apply the calculus of variations and optimal control theory to the problem of memristive switching. We have derived the optimal driving protocols for the following optimization problems: 
\begin{itemize}
    \item unconstrained switching of ideal memristors within a fixed interval of time (Subsec.~\ref{sec:2A1});
    \item unconstrained switching of ideal memristors within a variable interval of time (Subsec.~\ref{sec:2A3});
    \item unconstrained switching of memristive systems within a fixed interval of time (Subsec.~\ref{sec:2B1});
    \item switching of ideal memristors within a fixed interval of time in the presence of a constraint (Subsec.~\ref{sec:3A}).
\end{itemize}
An interesting finding is that, in unconstrained ideal memristor problems, the optimal trajectory corresponds to Joule losses occurring at a consistent rate (see Theorems 1 and 2 below). We compare our derived switching protocols to the cases of constant voltage or current. Our results show that the voltage
or current control can be used to reduce Joule losses in emerging ReRAM circuits and systems.

 The numerical simulations reported in this work were performed using Mathematica ver. 14.0.0.0. The comparison of optimal switching protocols is made with switching by constant voltage and current, which are included solely for comparative purposes.

This paper is structured as follows. 
Sec.~\ref{sec:2} is devoted to unconstrained optimization problems. Within Sec.~\ref{sec:2}, Subsec.~\ref{sec:2A} focuses on ideal memristors, while Subsec.~\ref{sec:2B} is dedicated to memristive systems. 
Constrained switching is discussed in Sec.~\ref{sec:3}.
Examples of optimal switching are given in Subsec.~\ref{sec:2A2}, Subsec.~\ref{sec:2A3}, Subsec.~\ref{sec:2B2}, and  Sec.~\ref{sec:3A2} where the calculus of variations and optimal control theory are applied to linear ideal memristors and memristive systems with a threshold. The paper concludes with a discussion of our results and future work.

\section{Unconstrained optimization problems} \label{sec:2} 

\subsection{Ideal memristors} \label{sec:2A}

\subsubsection{Minimization of Joule losses} \label{sec:2A1}

\begin{figure}[t]
\centering
\includegraphics[width=0.75\columnwidth]{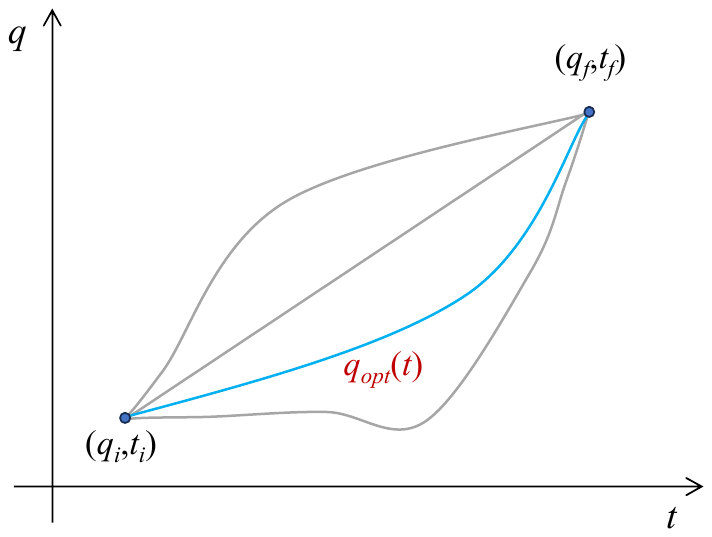}
\caption{Schematics of the optimal trajectory (blue curve labeled by $q_{opt}(t)$) and few non-optimal trajectories (grey curves) for the optimization problem in Subsec.~\ref{sec:2A1}. The optimal trajectory minimizes the Joule heat functional given by Eq.~(\ref{eq:1}).}
\label{fig:1}
\end{figure}

Let us first consider the problem of minimization of Joule losses in the switching of ideal memristors. 
Without loss of generality, the ideal memristors are described by Eq.~(\ref{eq:ideal}).
In this case, the energy cost to switch
from one state to another is a functional~\cite{arfken} with respect to the charge trajectory $q(t)$. This trajectory links the initial state of the memristor, $R_i$, at the initial moment of time, $t_i$, with its final state, $R_f$, at the final moment of time, $t=t_f$. For a given $q(t)$, the Joule heat, $Q[q(t)]$, is expressed by
\begin{equation}    Q[q(t)]=\int\limits_{(q_i,t_i)}^{(q_f,t_f)}\dot{q}^2R_M(q)\textnormal{d}t,
\label{eq:1}
\end{equation}
where $q_{i/f}$ is the initial/final charge ($R_M(q_{i/f})=R_{i/f}$), and 
it is assumed that $R_M(q)$ is known~\footnote{Eq.~(\ref{eq:1}) establishes a functional, which is an entity that produces a number when provided with a function.}. Our goal is to find the optimal trajectory, $q_{opt}(t)$, that minimizes the Joul heat functional, Eq.~(\ref{eq:1}), see the illustration in Fig.~\ref{fig:1}.

In principle, this problem is similar to the principle of least action in classical mechanics~\cite{goldstein2002classical}. Therefore, to determine the optimal trajectory we use the Euler-Lagrange equation that leads to the equation of motion  
\begin{equation}
\ddot{q}R_M(q)+\frac{1}{2}\dot{q}^2\frac{\textnormal{d}R_M(q)}{\textnormal{d}q}=0.
\label{eq:2}
\end{equation}
The first integral of Eq.~(\ref{eq:2}), corresponding to the energy conservation law in mechanics,
\begin{equation}
\dot{q}\sqrt{R_M(q)}=C_1,
\label{eq:3}
\end{equation}
can be integrated leading to
\begin{equation}
    \int \sqrt{R_M(q)}\textnormal{d}q=C_1\int\textnormal{d}t=C_1t+C_2, \label{eq:integral}
\end{equation}
where $C_1$ and $C_2$ are constants.

Note that the first integral, Eq.~(\ref{eq:3}), is nothing else than the 
conserved ``kinetic energy'', $R_M(q)\dot{q}^2$, for the Joule heat functional,  Eq.~(\ref{eq:1}). In the present context, it represents the power that is constant along the optimal trajectory. This observation is formulated in the following theorem.

\textbf{Theorem 1.} \textit{The optimal trajectory minimizing Joule losses in ideal memristors is characterized by constant power (in unconstrained problems).}

\subsubsection{Joule losses in linear memristors} \label{sec:2A2}

As an example of the above equations, consider the minimization of Joule losses in linear memristors.
Let us assume that in the region of interest, the memristance $R_M(q)$ can be approximated by a linear function, 
\begin{equation}
 R_M(q)=a+bq.   
 \label{Linear_memristor}
\end{equation}
Here, $a$ and $b$ are constants. In this case, the integral
in Eq.~(\ref{eq:integral}) can be easily evaluated and Eq.~(\ref{eq:integral}) is rewritten as
\begin{equation}
  \frac{2}{3b}\left(a+bq(t) \right)^{\frac{3}{2}}=C_1t+C_2.
\end{equation}
Next, the integration constants $C_1$ and $C_2$ are found using the initial and final point of the trajectory (see Eq.~(\ref{eq:1}) and Fig.~\ref{fig:1}). 
Explicitly, the charge trajectory minimizing the Joule losses is
\begin{eqnarray}
    q_{opt}(t)&=&\frac{1}{b}\left[ \frac{3b}{2}\left( C_1t+C_2\right)\right]^\frac{2}{3}-\frac{a}{b}  = \label{eq:8}\\
    &&
    =\frac{1}{b}\left[ \frac{R_i^\frac{3}{2}(t_f-t)+R_f^\frac{3}{2}(t-t_i)}{t_f-t_i} \right]^\frac{2}{3}-\frac{a}{b}.\nonumber
\end{eqnarray}
 The current corresponding to $q_{opt} (t)$ is
\begin{equation}
    I_{opt}(t)=
    \frac{2}{3b}
    \frac{R_f^\frac{3}{2}-R_i^\frac{3}{2}}{t_f-t_i}
    \left[ \frac{R_i^\frac{3}{2}(t_f-t)+R_f^\frac{3}{2}(t-t_i)}{t_f-t_i} \right]^{-\frac{1}{3}}
    \label{eq:9}
\end{equation}
and total Joule heat generated during the switching is
\begin{equation}
    Q_{opt}=
    \left(\frac{2}{3b}\right)^2
    \frac{\left(R_f^\frac{3}{2}-R_i^\frac{3}{2}\right)^2}{t_f-t_i}.    \label{eq:10}
\end{equation}
Using Eq.~(\ref{eq:3}) it is not difficult to verify that Theorem 1 is satisfied by $q_{opt} (t)$.

\begin{figure*}[tb]
\centering
(a)\;\;\includegraphics[width=0.8\columnwidth]{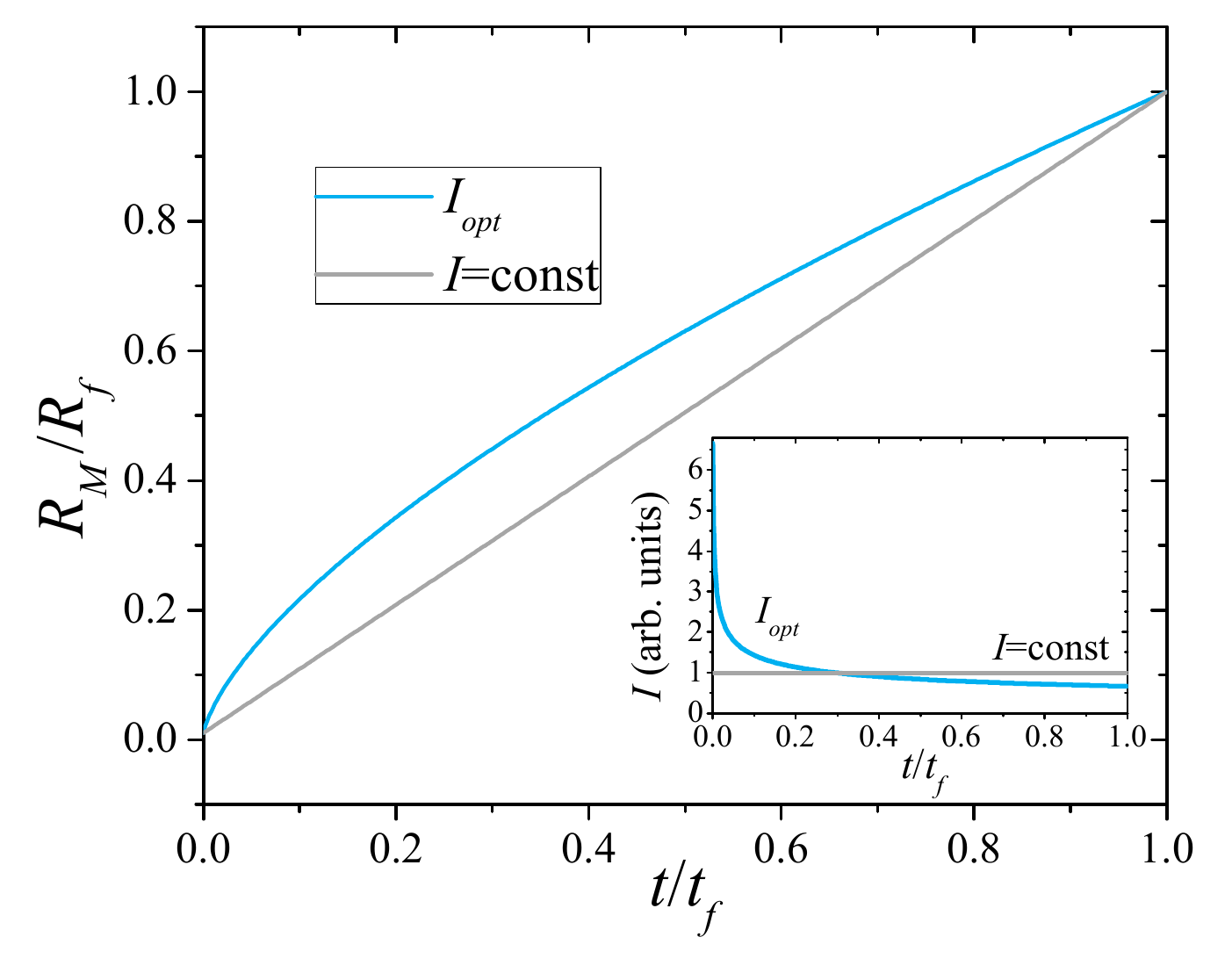} 
(b)\;\;\includegraphics[width=0.8\columnwidth]{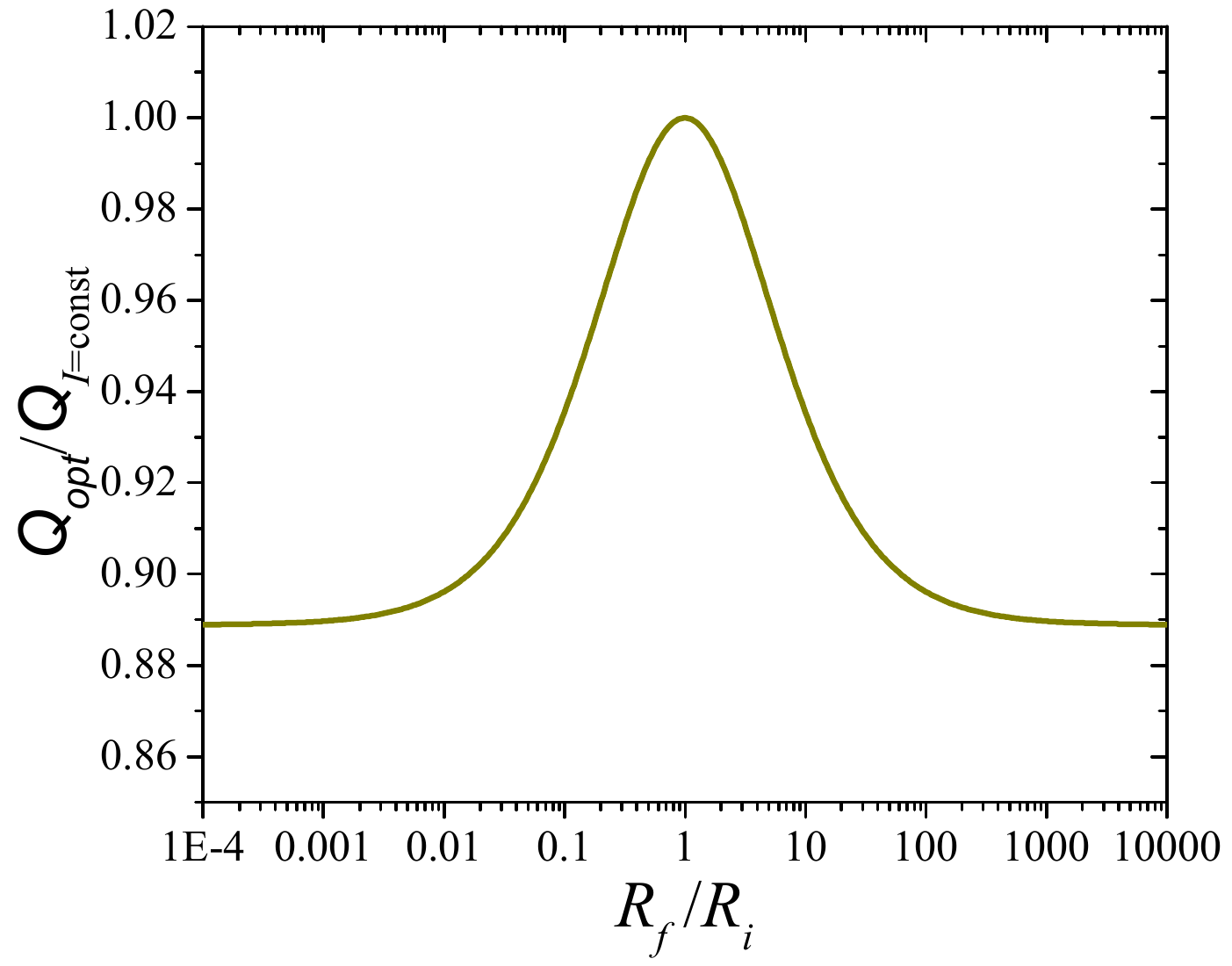}
\caption{Minimization of Joule losses in the ideal linear memristors (for the model see Eq.~(\ref{Linear_memristor})). (a) Memristance as a function of time for the cases of optimal current and constant current. The control currents are shown in the inset. This plot was obtained using $R_f/R_i=100$ and $t_i=0$. (b) Total Joule heat as a function of $R_f/R_i$. Asymptotically, $Q_{opt}/Q_{I(V)=\textnormal{const}}\rightarrow 8/9$ as $R_f/R_i\rightarrow 0$ or $\infty$.}
\label{fig:2}
\end{figure*}

It is interesting to compare $Q_{opt}$ for the optimal trajectory (Eq.~(\ref{eq:10})) to the Joule heat for the trajectory connecting $(q_i,t_i)$ to $(q_f,t_f)$ linearly,
\begin{equation}
    q_{I=\textnormal{const}}(t)=q_i\frac{t_f-t}{t_f-t_i}+q_f\frac{t-t_i}{t_f-t_i}.
\end{equation}
We note that this case corresponds to driving by a constant current.
Using Eq.~(\ref{eq:1}) one finds 
\begin{equation}
    Q_{I=\textnormal{const}}=
     \frac{1}{2b^2}
    \frac{(R_f^2-R_i^2)(R_f-R_i)}{t_f-t_i}.    \label{eq:12}
\end{equation}
It can be directly verified that the heat $Q_{I=\textnormal{const}}$ is always greater than the optimal Joule heat,  $Q_{I=\textnormal{const}}>Q_{opt}$, for all values $R_f\neq R_i$ and $t_f>t_i$  as it should be.

When the memristor is driven by a constant voltage $V$, the Ohm's law is simply the differential equation
$\dot{q}R_M(q)=V$ having the solution
\begin{equation}
    q_{V=\textnormal{const}}(t)=\frac{-a+\sqrt{R_i^2+2bV(t-t_i)}}{b}.
\end{equation}
Interestingly, in this case the total Joule heat is the same as in the case of constant current, Eq.~(\ref{eq:12}). As the voltage as a function of switching time interval can be written as
\begin{equation}
V=\frac{1}{2b}\frac{R_f^2-R_i^2}{t_f-t_i},
\end{equation}
another expression for $Q_{V=\textnormal{const}}$ is
\begin{equation}
 Q_{V=\textnormal{const}}=\frac{V(R_f-R_i)}{b}.
\end{equation}

Certain results found for linear memristors are shown graphically in Fig.~\ref{fig:2}. In particular, switching linear memristors using optimal control reduces Joule losses by up to a factor of $8/9$ or approximately 11\% compared to switching with constant voltage or current. Figure~\ref{fig:2}(a) shows that reducing Joule losses is achieved by increasing the current $I$ when the resistance $R_M$ is smaller, and vice versa.

\subsubsection{Simultaneous minimization of Joule losses and switching time} \label{sec:2A3}

Equations (\ref{eq:10}) and (\ref{eq:12}) suggest that Joule losses can be minimized by increasing the switching time. 
However, in numerous applications, a high operating frequency is essential. In this part of the paper, we explore how to reconcile the conflicting demands of low losses and rapid switching.

Thus, we would like to find the trajectory that minimizes Joule losses and switching time simultaneously. For this purpose, the cost functional can be selected as
\begin{equation}
F=f(Q[q(t)])+g(t_f-t_i),
\end{equation}
where  $f(x)$ and  $g(x)$ are monotonously increasing functions, and $Q[q(t)]$ is given by Eq.~(\ref{eq:1}). In the linear case, the cost functional is simply
\begin{equation}
F=w_1Q[q(t)]+w_2\cdot (t_f-t_i),  \label{eq:14}
\end{equation}
where $w_i$ are the positive weights (distinct units).  

To find the minimum value of the cost functional, we independently vary $q(t)$ and $t_f$. The variation of $q(t)$ gives the same Eq.~(\ref{eq:2}), and the variation of $t_f$ gives the additional equation
\begin{equation}
\left.\left(w_2-w_1 \dot{q}^2R_M(q)\right)\right|_{t=t_f}=0,  \label{eq:15}
\end{equation}
which, according to  Eq.~(\ref{eq:3}), can be simplified to $w_2=w_1 C_1^2$. As the first integral of Eq.~(\ref{eq:2}) is given by the same Eq.~(\ref{eq:3}), the power is constant along the optimal trajectory. Therefore, we state the following theorem.

\textbf{Theorem 2.} \textit{The optimal trajectory minimizing Joule losses and switching time in ideal memristors is characterized by constant power (in unconstrained problems).}

Returning to the example in Subsec.~\ref{sec:2A2}, we can use Eq.~(\ref{eq:10}) to express the power in Eq.~(\ref{eq:15}) and eventually obtain
the optimal time
\begin{equation}
    (t_f-t_i)_{opt}=\sqrt{\frac{w_1}{w_2}}\frac{2}{3b}\left(R_f^\frac{3}{2}-R_i^\frac{3}{2}\right).
\end{equation}
At this value of $t_f-t_i$,
\begin{equation}
    Q_{opt}=\sqrt{\frac{w_2}{w_1}}\frac{2}{3b}\left(R_f^\frac{3}{2}-R_i^\frac{3}{2}\right).
\end{equation}

\subsection{Memristive systems} \label{sec:2B}

\subsubsection{Lagrangian function} \label{sec:2B1}

Next, we consider a voltage-controlled memristive device satisfying Eqs.~(\ref{eq1})-(\ref{eq2}). In the following discussion, we focus on a scenario with a single internal state variable. However, our primary equations can be adapted to more complex situations.

Our aim is to identify the most efficient driving protocol that minimizes Joule losses when the device switches from $x(t_i)=x_i$ to $x(t_f)=x_f$. Mathematically, we consider a model with the Joule heat functional defined by
\begin{equation}
Q=\int\limits_{t_i}^{t_f}\frac{V^2(t)}{R_M(x)}\textnormal{d}t \;\; \rightarrow \;\; \textnormal{min},
\label{eq:heat}
\end{equation}
where $x$ is the function of time according to Eq. (\ref{eq2}). 

According to the general scheme~\cite{alekseev2013optimal}, the Lagrangian function is written as
\begin{equation}
\mathcal{L}=\lambda_0\int\limits_{t_i}^{t_f}\frac{V^2(t)}{R_M(x)}\textnormal{d}t+\int\limits_{t_i}^{t_f}\Lambda(t)\left[ \dot x(t)-f(x,V)\right]\textnormal{d}t+l,
\label{funct_Lagr}
\end{equation}
where $\lambda_0$ and $l$ are constants, and $\Lambda(t)$ is a function of time known as the Lagrange multiplier. From (\ref{funct_Lagr}), the Lagrangian is 
\begin{equation}
    L(x,\dot x, V,\Lambda)=\lambda_0\frac{V^2(t)}{R_M(x)}+\Lambda(t)\left[ \dot x(t)-f(x,V)\right].
\end{equation}
The necessary conditions for an extrema are the following~\cite{alekseev2013optimal}:
\begin{itemize}
    \item Euler-Lagrange equation: $\textnormal{d}\left( \partial L/\partial \dot x \right) /\textnormal{d} t= \partial L/\partial x $;
    \item Stationary condition with respect to $V$: $\partial L/\partial V=0$;
    \item Transversality conditions $L'_{\dot x}(t_i)=l'_{x(t_i)}$ and $L'_{\dot x}(t_f)=-l'_{x(t_f)}$, where $l=\lambda_i x(t_i)+\lambda_fx(t_f)$ is the terminant.
\end{itemize}

The last condition leads to $\Lambda(t_i)=\lambda_i$ and $\Lambda(t_f)=-\lambda_f$. As $\lambda_i$ and $\lambda_f$ are arbitrary constants, these conditions can be omitted. Therefore, our optimization problem is defined by the following set of equations:
\begin{eqnarray}
\Lambda(t)&=&\lambda_0\frac{2 V(t)}{R_M(x)f'_{V}(x,V)}, \label{eq:opt1}\\
\frac{\textnormal{d} \Lambda(t)}{\textnormal{d} t}&=&\lambda_0V^2(t)\frac{\textnormal{d}}{\textnormal{d} x}\left[ \frac{1}{R_M(x)} \right]-\Lambda (t)f'_x(x,V), \label{eq:opt2}\\
\frac{\textnormal{d} x(t)}{\textnormal{d} t}&=&f(x,V), \label{eq:opt3}\\
x(t_i) &=& x_i \; , \; x(t_f) = x_f. \label{eq:opt4}
\end{eqnarray}

\begin{figure*}[tb]
\centering
(a)\;\;\includegraphics[width=0.8\columnwidth]{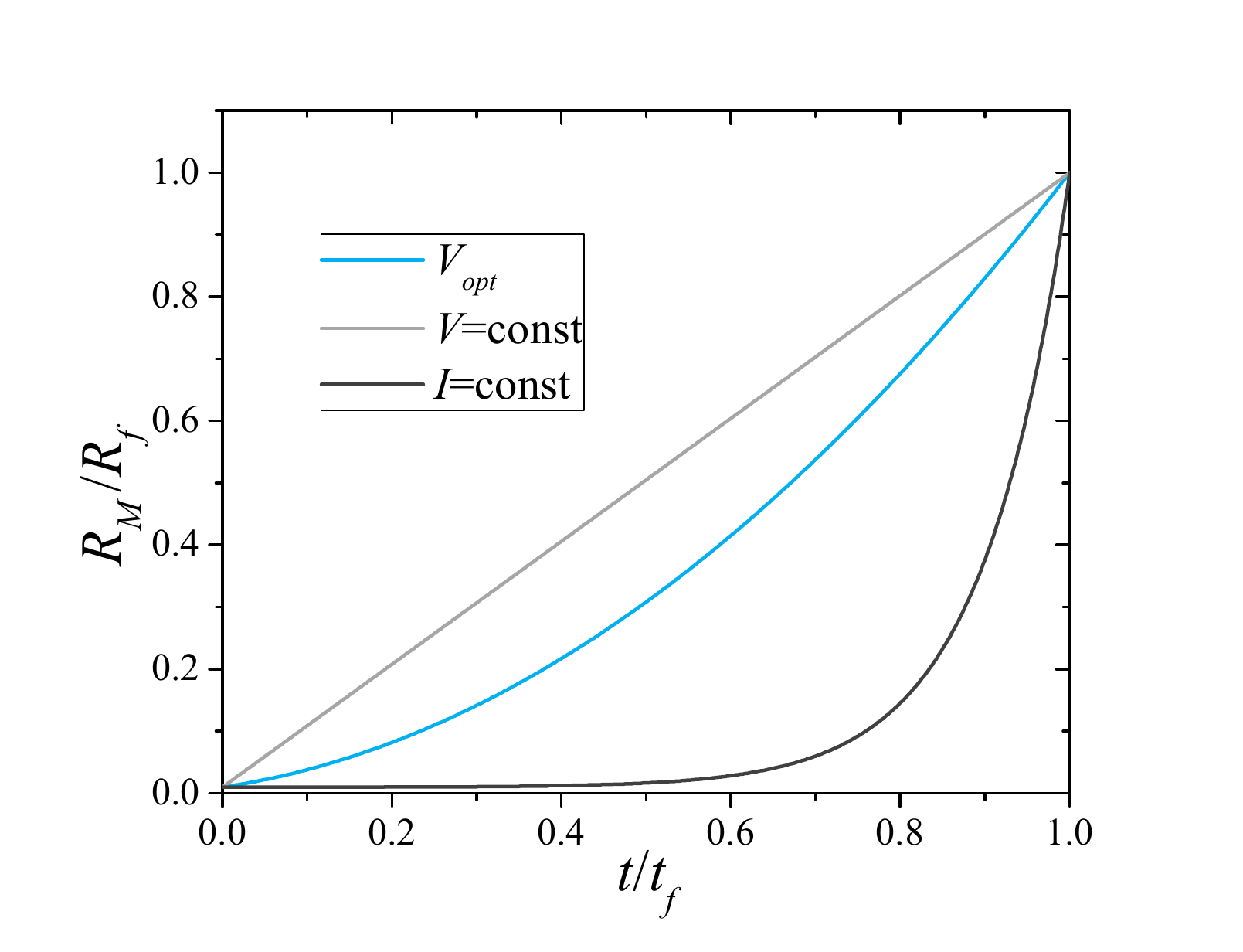}
(b)\;\;\includegraphics[width=0.8\columnwidth]{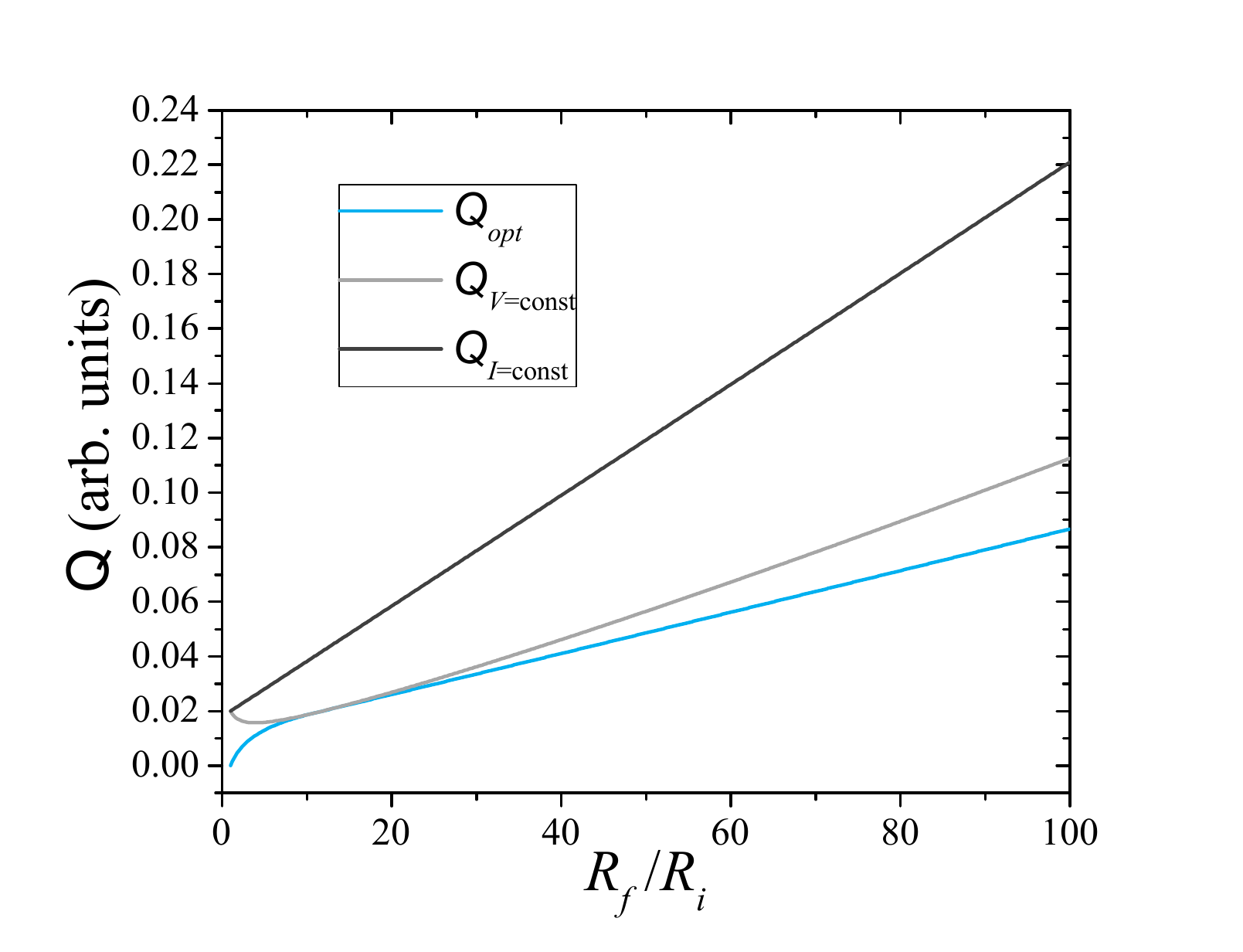}
\caption{Minimization of Joule losses in memristive devices. (a) Memristance as a function of time for the cases of optimal control, constant voltage control, and constant current control. (b) Joule heat as a function of $R_f/R_i$. These figures were obtained using the following set of parameter values:  $R_{on}=1$~k$\Omega$, $R_{off}=100$~k$\Omega$, $k=0.5$~$(\textnormal{V}\cdot\mu\textnormal{s})^{-1}$, 
$V_{off}=0.1$~V,
$t_i=0$, $R_i=R_{on}$, $t_f=2$~$\mu$s, and (a) $R_f=R_{off}$, (b) $R_{on}\leq R_f\leq R_{off}$.}
\label{fig:3}
\end{figure*}

\subsubsection{Derivation of ideal memristor equations} \label{sec:2B2}

Here, we show that Eq.~(\ref{eq:2}) for ideal memristors in Subsec.~\ref{sec:2A1} can be derived directly from Eqs.~(\ref{eq:opt1})-(\ref{eq:opt4}). 
For this purpose, using $f(x,V)=V/R_M(x)$ we first obtain the partial derivatives $f'_V=1/R_M(x)$ and $f'_x=-R_M'(x)V/R_M^2(x)$. Substituting these partial derivatives into Eq. (\ref{eq:opt1}) yields
\begin{equation}
\Lambda(t)=2\lambda_0\frac{V(t)}{R_M(x)/R_M(x)}=2\lambda_0V(t). \label{eq:17}
\end{equation}
Using Eq.~(\ref{eq:17}), Eq.~(\ref{eq:opt2}) can be rewritten as
\begin{equation}
    \dot V(t)=\frac{1}{2} V^2(t) \frac{R_M'(x)}{R_M^2(x)}. \label{eq:18}
\end{equation}
Taking into account that $V(t)=R_M(q)\dot q$, Eq.~(\ref{eq:18}) leads to
\begin{equation}
    \ddot q R_M(q)+\frac{1}{2}\left( \dot q\right)^2\frac{\textnormal{d} R_M}{\textnormal{d}q}=0,
\end{equation}
which is precisely Eq.~(\ref{eq:2}) in Subsec.~\ref{sec:2A1}. Therefore, the results presented in Subsecs.~\ref{sec:2A1} and \ref{sec:2A1} remain consistent when applying our comprehensive optimization theory formulated for the memristive devices in the current subsection.

\subsubsection{Optimal control of a threshold-type memristive device} \label{2B3}

Traditional experimental memristive devices~\cite{Song2023-wo,holy} demonstrate switching with a threshold, which is essential for nonvolatile information storage. In the following, we apply the optimization approach from Subsec.~\ref{sec:2B1} to a memristive device described by a model featuring a threshold. Specifically, we consider a device described by the equations~\cite{pershin09b}
\begin{eqnarray}
    R_M(x)&=&R_{on}+x(R_{off}-R_{on}) , \label{eq:Rx} \\
    \frac{\textnormal{d}x}{\textnormal{d}t}&=&
     \begin{cases}
k(V-V_{off}),  & V>V_{off}>0,\\
0, & V_{on}\leq V \leq V_{off},\\
k(V-V_{on}), & V<V_{on}<0, \label{eq:threshold}
     \end{cases}
\end{eqnarray}
where $k>0$ is the constant, $R_{on}$ and $R_{off}$ are the on-state and off-state resistances, and $V_{on}$ and $V_{off}$ are the thresholds. According to the above equations, when the voltage is greater than $V_{off}$, the memristive system switches to the high-resistance state, $R_{off}$. 

For the sake of definiteness, let us consider the transition $x_i\rightarrow x_f$ {\it assuming} that $V>V_{off}$. As $f(x,V)=k(V-V_{off})$, Eqs.~(\ref{eq:opt1})-(\ref{eq:opt3}) take the form
\begin{eqnarray}
\Lambda(t)&=&2\lambda_0\frac{V(t)}{kR_M(x)}, \label{eq:optTr1}\\
\frac{\textnormal{d} \Lambda(t)}{\textnormal{d} t}&=&-\lambda_0V^2(t)\frac{R_M'(x)}{R_M^2(x)}, \label{eq:optTr2}\\
\frac{\textnormal{d} x(t)}{\textnormal{d} t}&=&k(V-V_{off}). \label{eq:optTr3}
\end{eqnarray}
Next, without the loss of generality, we set $\lambda_0=1$. Substituting $V(t)/R_M(x)$ from Eq.~(\ref{eq:optTr1}) into Eq.~(\ref{eq:optTr2}) we arrive at
\begin{equation}
    \frac{\textnormal{d} \Lambda(t)}{\textnormal{d} t}=-R_M'(x)\left(\frac{k}{2} \right)^2\Lambda^2(t). \label{eq:40}
\end{equation}
The solution of Eq.~(\ref{eq:40}) can be expressed as
\begin{equation}
    \Lambda(t)=\frac{1}{\frac{k^2}{4}R_M'(x)t+C_1} \label{eq:28},
\end{equation}
where $C_1$ is the integration constant. 

Using (\ref{eq:optTr1}) we obtain the following expression  for the voltage
\begin{equation}
    V(t)=\frac{k}{2}\frac{R_M(x)}{\frac{k^2}{4}R_M'(x)t+C_1} \label{eq:29}
\end{equation}
that can be further simplified using the linear dependence of $R_M$ on $x$, Eq.~(\ref{eq:Rx}). The substitution of (\ref{eq:29}) and (\ref{eq:Rx}) into 
the state equation (\ref{eq:optTr3}) leads to
\begin{equation}
    \frac{\textnormal{d}x}{\textnormal{d}t}=\frac{2R_{on}}{R_M'(x)(t+t_0)}-kV_{off}+\frac{2}{t+t_0}x, \label{eq:x}
\end{equation}
where $t_0=4C_1/(k^2R_M'(x))$ is employed for compactness in place of $C_1$.

The solution of Eq.~(\ref{eq:x}) can be written as 
\begin{equation}
    x(t)=C(t+t_0)^2+kV_{off}(t+t_0)-\frac{R_{on}}{R_M'(x)},
\end{equation}
where $C$ is an arbitrary constant. Consequently,
\begin{equation}
    R_M(t)=CR_M'(x)(t+t_0)^2+kR_M'(x)V_{off}(t+t_0).
    \label{eq:45}
\end{equation}
The values of the arbitrary constants $C$ and $t_0$ can be determined by setting $R_M(t_i)$ equal to $R_i$ and $R_M(t_f)$ equal to $R_f$.

The voltage as a function of time is found by utilizing Eq.~(\ref{eq:29}), which yields 
\begin{eqnarray} 
V(t)&=&2V_{off}+\frac{2C}{k}(t+t_0), \label{eq:47} \\
I(t)&=&\frac{2}{kR_M'(x)}\frac{1}{t+t_0}.  \label{eq:47_}
\end{eqnarray} 
According to Eq.~(\ref{eq:47}) the control voltage is a linear function of time.

Using the initial and final resistances, one can obtain
\begin{equation}
\label{C}
    C=\frac{R_f+R_i-\sqrt{4R_iR_f+[kV_{off}R_M'(x)(t_f-t_i)]^2}}{R_M'(x)(t_f-t_i)^2}
\end{equation}
and
\begin{equation}
    t_0=\frac{1}{2C}\left\{\frac{R_f-R_i}{(t_f-t_i)R_M'(x)}-kV_{off}\right\}-\frac{t_f+t_i}{2}.
    \label{eq:t0}
\end{equation}

It should be noted that the solution involving the positive root in Eq.~(\ref{C}) has been discarded, as it does not meet the condition $V(t)>V_{off}$, which we have assumed from the beginning of this subsection. Furthermore, the sum $t_0+t_i>0$, since we are operating under the assumption that $R_f>R_i$ in all cases. That is important because it guarantees the finiteness of the switching current $I(t)$ defined by Eq.~(\ref{eq:47_}) for all moments of time. 

It is also observed that $C$ is positive if $R_f> R_i + kV_{off}R_M'(x)(t_f-t_i)$. According to Eq.~(\ref{eq:47}), this implies that the optimal control voltage rises over time, surpassing the double threshold voltage $2V_{off}$ even at the initial moment $t_i$. When $R_f = R_i + kV_{off}R_M'(x)(t_f - t_i)$, the parameter $C$ becomes zero, and as per Eq.~(\ref{eq:47}), the optimal control voltage remains constant at $V(t) = 2V_{off}$ throughout the entire switching interval, from $t_i$ to $t_f$.

In the parameter domain  $R_f< R_i+ kV_{off}R_M'(x)(t_f-t_i)$, the solution given by Eqs.~(\ref{eq:45})-(\ref{eq:t0}), ceases to be optimal. In addition, this solution does not even satisfy the assumption that $V(t_f)>V_{off}$, when $R_f\rightarrow R_i$. It is clear that in this case, the optimal strategy is to quickly switch the memristor from the initial state $R_i$ to the final state $R_f$ during a finite time interval $\Delta t$ by applying the voltage exceeding the threshold voltage $V_{off}$ and then applying zero voltage for the remaining time. It turns out that the magnitude of optimal voltage is the same as in the case when $R_f= R_i+ kV_{off}R_M'(x)(t_f-t_i)$,
\begin{eqnarray}
   V(t)&=&
     \begin{cases}
2V_{off},  & t_i<t<t_i+\Delta t,\\
0, & t_i+\Delta t<t<t_f,
     \end{cases}
\end{eqnarray}
while in accordance with Eqs.~(\ref{eq:Rx})-(\ref{eq:threshold}) memristance is a piecewise linear function of time
\begin{eqnarray}
   R(t)=
     \begin{cases}
R_i+kV_{off}R_M'(x)(t-t_i),  & t_i<t<t_i+\Delta t,\\
R_f, & t_i+\Delta t<t<t_f,
     \end{cases}
\end{eqnarray}
where $\Delta t=(R_f-R_i)/(kR_M'(x)V_{off})$.


 Finally, the minimal Joule heat  is expressed by
\begin{equation}
    Q_{opt}=\frac{4}{k^2R_M'(x)}\left[ C\left( t_f-t_i \right) + kV_{off}\ln\left(\frac{t_f+t_0}{t_i+t_0}\right)\right],
\end{equation}
when $R_f-R_i\geq kV_{off}R_M'(x)(t_f-t_i)$,\\ and
\begin{equation}
    Q_{opt}=\frac{4V_{off}}{kR_M'(x)}\ln\frac{R_f}{R_i},
\end{equation}
when $R_f-R_i\leq kV_{off}R_M'(x)(t_f-t_i)$.

It is interesting to compare these minimal Joule losses with the Joule losses for other possible switching regimes: switching at constant voltage $V=V_0$ and constant current $I=I_0$. 
For the constant voltage switching protocol by integrating in Eq. ~(\ref{eq:heat}) over time
with Eq.~(\ref{eq:Rx}) and Eq.~(\ref{eq:threshold}) taking into account, we get
\begin{equation}
    Q_{V=\textnormal{const}}=V_0^2\frac{t_f-t_i}{R_f-R_i}\ln\frac{R_f}{R_i},
    \label{eq:Q_Vconst}
\end{equation}
where $V_0=V_{off}+(R_f-R_i)/(kR_M'(x)(t_f-t_i))$.

Similarly, for the constant current switching protocol, we find for the Joule losses
\begin{equation}
    Q_{I=\textnormal{const}}=I_0\left[V_{off}(t_f-t_i)+
    \frac{R_f-R_i}{kR_M'(x)}
    \right],
    \label{eq:Q_Iconst}
\end{equation}
where the constant current $I_0$ is  a unique solution of the following equation
\begin{equation}
   e^{kR_M'(x)(t_f-t_i)I_0}=\frac{R_fI_0-V_{off}}{R_iI_0-V_{off}},
    \label{eq:Iconst}
\end{equation}
in the domain $I_0>V_{off}/R_i$, and the voltage depends exponentially on time as
\begin{equation}
   V(t)=V_{off}+(R_iI_0-V_{off})e^{kR_M'(x)I_0(t-t_i)}.
\end{equation}

Fig.~\ref{fig:3}(a) exemplifies the time dependence of the memristance for several driving protocols. We note that in the cases of optimal control, constant voltage control,
and constant current control, the memristance increases quadratically (Eq.~(\ref{eq:45})), linearly, and exponentially with time, respectively. Based on the information presented in Fig.~\ref{fig:3}(b), the use of optimal control has the potential to decrease Joule losses by approximately 27\% compared to switching based on constant voltage and by approximately 35\% in contrast to switching based on constant current. Qualitatively, the superior efficiency of constant voltage switching relative to constant current switching can be attributed to the dynamics in the constant voltage case that is closer to the optimal one. In particular, this can be recognized in Fig.~\ref{fig:3}(a).

\section{Constrained optimization problems} \label{sec:3}

\subsection{Ideal memristors} \label{sec:3A}

\subsubsection{Pontryagin’s principle} \label{sec:3A1}
 
Let us now address a more complex case involving the optimal control of memristive switching while taking into account inequality constraints. Specifically, we will focus on the case of an ideal memristor with a current constraint, where we assume that the current $I(t)$ passing through the device is limited by a critical (compliance) current $I_c$, such that $| I(t)| \leq I_c$. Consequently, we are interested in minimizing Joule losses, which can be expressed through the functional~(\ref{eq:1}), while also imposing the current limiting constraint $| \dot{q} | \leq I_c$.

\begin{figure*}[tb]
\centering
(a)\includegraphics[width=0.8\columnwidth]{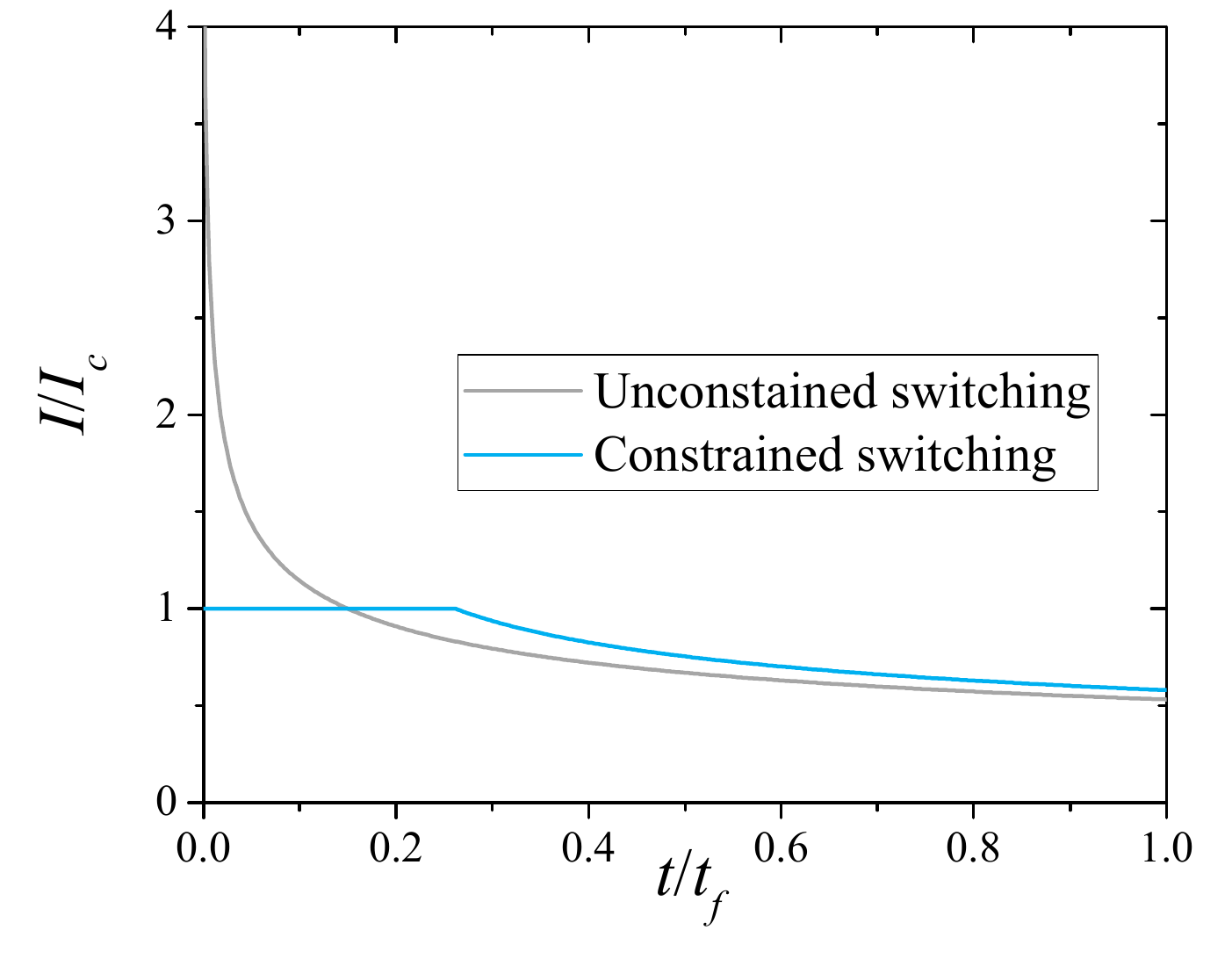} 
(b)\includegraphics[width=0.8\columnwidth]{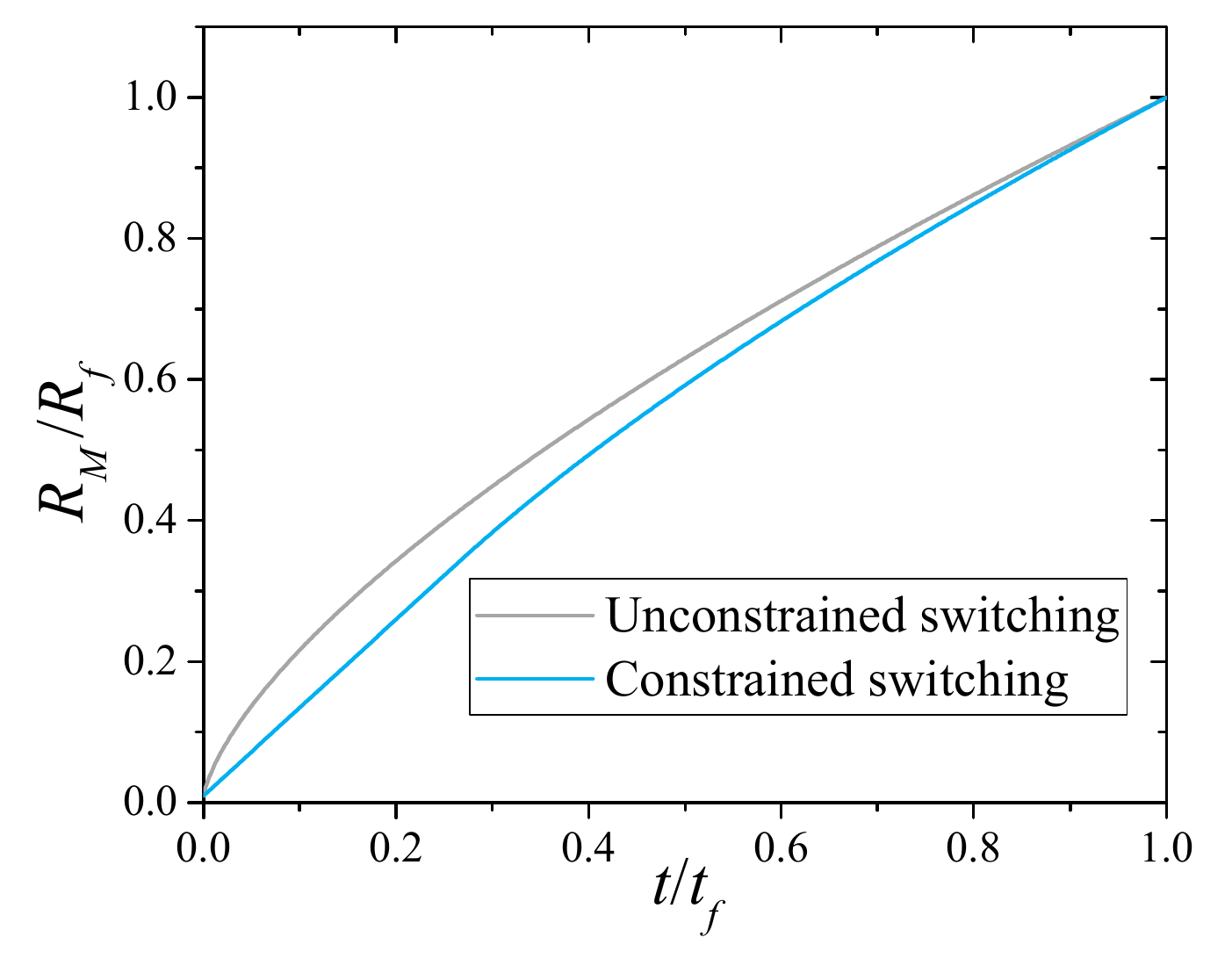} 
\caption{Constrained versus unconstrained control of memristive switching. (a) Control current as a function of time. (b) Memristance as a function of time. This figure was obtained using the following set of parameter values: 
$R_{on}=R_i=1$~k$\Omega$, $R_{off}=R_f=100$~k$\Omega$, 
$I_c=0.25$~mA, 
$b=10^2$~$k\Omega/(\textnormal{mA}\cdot\mu\textnormal{s})$, 
$t_i=0$ and $t_f=5$~$\mu$s. 
}
\label{fig:4}
\end{figure*}

It is evident that when the critical current $I_c$ is sufficiently large, the solution of the problem is determined by the same Euler-Lagrange equation (Eq.~(\ref{eq:2})). The situation becomes more complicated when the current, calculated using the solution derived from Eq.~(\ref{eq:integral}) with specified boundary conditions, exceeds the critical current at least at a certain point in time. In such a situation, it becomes necessary to incorporate Pontryagin's principle of maximum (minimum) to determine the optimal control $q(t)$. To facilitate this, it is convenient to introduce a new variable $u$ following the definition~\footnote{It is evident that the definition of $u$ is identical to the current. Nevertheless, a distinct notation is required since, in Pontryagin's principle, $u$ is treated as an independent variable.} 
\begin{equation}
u=\dot q.
\label{eq:u_def}
\end{equation}
This way we transfer the constraints into the configuration space
\begin{equation}
    |u|\leq I_c.
    \label{eq:restrict}
\end{equation}

 The Lagrangian function for this optimal problem is written as (see~\cite{alekseev2013optimal})
\begin{equation}
\mathcal{L}=\int\limits_{t_i}^{t_f}
\left[\lambda_0 R_M(q)u^2+p(t)\left( \dot q(t)-u\right)\right]\textnormal{d}t+l,
\label{funct_Lagr2}
\end{equation}
where constant $\lambda_0\geq 0$ and function $p(t)$ are the unknown Lagrange multipliers.  In the subsequent discussion, we omit the term $l=\lambda_i q(t_i)+\lambda_f q(t_f)$ from the Lagrangian function~(\ref{funct_Lagr2}) since the transversality conditions associated with it do not impose any limitations on the solution of the problem under consideration. 

The optimal control must satisfy not only Euler-Lagrange equation with respect to $q$,
\begin{equation}
    \lambda_0 R_M'(q)u^2-\frac{\textnormal{d}p(t)}{\textnormal{d}t}=0,
    \label{Euler-Lagrange2}
\end{equation}
but also the Pontryagin's principle of maximum (minimum) with respect to $u$,
\begin{equation}
    \min_{|\eta|\leq I_c}\left[\lambda_0 R_M(q(t))\eta^2-p(t)\eta\right]=
    \lambda_0 R_M(q(t))u^2-p(t)u.
    \label{pontryagin}
\end{equation}
The last condition implies that, at any instant $t$, the optimal control $u(t)$ minimizes the expression in Eq.~(\ref{funct_Lagr2}) within the range $-I_c\leq u\leq I_c$ for the optimal solution $q(t)$ and $p(t)$.

Let us first examine the specific case of $\lambda_0=0$. Considering the Euler-Lagrange equation~(\ref{Euler-Lagrange2}), it follows that $p(t)$ is constant. Applying Pontryagin's principle~(\ref{pontryagin}) with $\lambda_0=0$, we can determine the optimal control for the function $u(t)=I_c\textnormal{sign } p(t)=\pm I_c$. Therefore, in the case where $\lambda_0=0$, the constant maximum current $\dot q=u=\pm I_c$ is maintained throughout the entire duration of the memristor switching.

This special solution is only possible under certain boundary conditions, specifically when $(q_f-q_i)/(t_f-t_i)=\pm I_c$. If this condition is not satisfied,  this particular solution is not viable, and $\lambda_0>0$. Furthermore, we can assume $\lambda_0=1$ in the following discussion without loss of generality.

The minimization of the quadratic function in the LHS of Eq.~(\ref{pontryagin}) gives the following optimal control for the function $u(t)$:
\begin{eqnarray}
\label{eq:u_opt} 
    u(t)&=&
     \begin{cases}
\frac{p(t)}{2R_M(q)},  & \textnormal{if } \frac{|p(t)|}{2R_M(q)} < I_c,\\
I_c\textnormal{sign } p(t), & \textnormal{otherwise}.
     \end{cases}\; .
\end{eqnarray}

The system of equations~(\ref{eq:u_def}), (\ref{Euler-Lagrange2}), with $\lambda_0=1$, and (\ref{eq:u_opt}) fully determines the optimal switching of the ideal memristor. It is evident that the current $u(t)$ satisfies the constraints specified in Eq.~(\ref{eq:restrict}). Furthermore, by eliminating the Lagrange multiplier $p(t)$ from Euler-Lagrange equation (Eq.~(\ref{Euler-Lagrange2})) using Eq.~(\ref{eq:u_opt}) for $u(t)<I_c$, we obtain the optimal trajectory for the ideal memristor without constraints, as expected. Therefore, it is apparent that achieving optimal control for the ideal memristor under current restrictions generally involves a smooth and continuous connection of the solution characterized by constant power (refer to Eq.~(\ref{eq:integral})) with the solution represented by the maximum possible current $q=\pm I_c t+C_3$, where $C_3$ is an arbitrary constant. 

\subsubsection{Joule losses in linear memristors} \label{sec:3A2}

We will illustrate the method discussed above by considering the switching of a linear memristor, considered previously in Subsec.~\ref{sec:2A2}. However, now we also assume that the maximum possible current is limited by the critical (compliance) value $I_c$. It is also assumed that during the time period from $t_i=0$ to $t_f$, the memristance changes from $R_M(0)=R_i$ to $R_M(t_f)=R_f>R_i$. Furthermore, we consider the most interesting case where at the initial moment of time the optimal unrestricted current $I_{opt}(t=t_i)$, given by Eq.~(\ref{eq:9}),
 greater than $I_c$. This corresponds to the following inequality:
 \begin{equation}
 \label{ineq}
     2 \frac{R_f^{3/2}-R_i^{3/2}}{3bt_f\sqrt{R_i}}>I_c.
 \end{equation}

Following the previous discussion, to minimize Joule heat, it is necessary to apply the highest current $I_c$ for a duration of up to $t_c$. In the case of the linear memristor described by Eq.~(\ref{Linear_memristor}), this leads to the following resistance change:
\begin{equation}
    R_M(t)=R_i+bI_c t,  \;\;\; \textnormal{for }0\leq t\leq t_c.
    \label{eq:59}
\end{equation}
Starting at $t=t_c$,  the optimal solution is given by Eq.~(\ref{eq:8}):
\begin{equation}
    R_M(t)=
    \left[ \frac{R_c^\frac{3}{2}(t_f-t)+R_f^\frac{3}{2}(t-t_c)}{t_f-t_c} \right]^{\frac{2}{3}},\;\;
    \textnormal{for }t_c\leq t\leq t_f,
    \label{eq:60}
\end{equation}
where resistance $R_c=R_M(t_c)$ calculated with the use of Eq.~(\ref{eq:59}).

The values of the parameters $t_c$ and $R_c$ need to be determined from the continuity condition of the memristance, $R_M(t)$, and the current, $I(t)=\dot R_M/b$, at the instant $t_c$ when the control change occurs. This leads to the following equations for $R_c$ and $t_c$:
\begin{equation}
    \sqrt{R_c^3}-3(R_i+bI_ct_f)\sqrt{R_c}+2\sqrt{R_f^3}=0,
    \label{RcEq}
\end{equation}
and
\begin{equation}
    t_c=\frac{R_c-R_i}{bI_c}.
    \label{tcEq}
\end{equation}

Note that an obvious condition must be met
for the final state, $R_f$, to be reachable from the initial state, $R_i$, during time interval $t_f$ due to the existance of the maximum possible current $I_c$:
 \begin{equation}
    R_f<R_i+bI_c t_f \;.
    \label{ineq2}
\end{equation}
This inequality guarantees the existence of two positive roots of Eq.~(\ref{RcEq}), the smallest of which is smaller than $R_f$, while the other is larger. This smallest positive root of Eq.~(\ref{RcEq}) determines the resistance $R_c$ and moment of time $t_c$ (see Eq.~(\ref{tcEq})), where the optimal control changes. Note that the inequality~(\ref{ineq}) guarantees that $R_c>R_i$.

Fig.~\ref{fig:4} illustrates our results on the optimal control of an ideal linear memristor with a current constraint. 
Fig.~\ref{fig:4}b represents memristance as a function of time calculated by using Eqs.~(\ref{eq:59}), 
(\ref{eq:60}) for the case of constrained switching and by using Eqs.~(\ref{Linear_memristor}), 
(\ref{eq:8}) for the case of unconstrained switching. 
The corresponding currents, which are proportional to the derivatives of the memristances with respect to time, are presented in Fig.~\ref{fig:4}a. Clearly, for the used set of parameters, inequalities~(\ref{ineq}) and (\ref{ineq2}) are satisfied. Thus, it allows us to find the only solution of Eqs.~(\ref{RcEq}) and ~(\ref{tcEq}), $R_c/R_f=0.34$ and $t_c/t_f=0.26$, which determines the resistance and time moment of control change for this set of parameters.

\subsection{Memristive systems} \label{sec:3B}
In principle, the general approach developed for memristive systems in Subsec.~\ref{sec:2B} combined with the Pontryagin principle (which we used for the analysis of the switching of ideal memristors with a current constraint in Sect.~\ref{sec:3A}) can be employed to optimize the switching of memristive devices (Eqs.~(\ref{eq1}) and (\ref{eq2})) in the presence of constraints. However, our derivations in Subsec.~\ref{sec:3A} suggest that optimal switching protocols, in fact,  can be obtained using a simplified approach, at least in certain cases. Such cases include (but are not limited to) some simple (first-order) memristive models and unidirectional switching. 

Consider, for example, situations that involve current compliance (see Eq.~(\ref{eq:restrict})). The idea is to link the two operational conditions, specifically, with and without current compliance. First of all, one can consider the unconstrained scenario (using the approach outlined in Subsec.~\ref{sec:2B}). If Eq.~(\ref{eq:restrict}) holds consistently, then the solutions for both constrained and unconstrained problems coincide. Otherwise, the solutions with and without the constraint should be combined while adhering to the criteria of (i) maintaining the continuity of the control parameter and (ii) minimizing Joule losses. 

\section{Discussion}

In this theoretical study, novel strategies have been devised to reduce Joule losses that arise when memristive devices are switched. Various hypothetical scenarios have been examined, and substantial energy savings have been shown to be possible by following specific driving protocols proposed in this research. 

It is important to note that our findings are model-dependent, implying that no universal protocol can be applied to all memristive devices. To implement our method(s), first, the device model must be identified. Subsequently, our optimization techniques can be utilized to determine the optimal driving protocol for that specific device model. The final step involves verification. Based on our experience, it is always beneficial to compare the derived protocol with other easily verifiable scenarios, such as switching by constant voltage or current. This comparison is crucial because the solution to the Euler-Lagrange equation represents an extremum, either a minimum or a maximum.

One might wonder about the experimental significance of the findings reported in this article, and indeed they are relevant. Our understanding of ideal devices (Subsec.~\ref{sec:2A}) indicates that their optimal switching involves constant power. In threshold-type memristive systems (Subsec.~\ref{2B3}), although constant power conditions do not strictly apply, a similar pattern remains: lower voltages are paired with lower resistances and vice-versa. These observations could be utilized in an empirical optimization.

Furthermore, the model described by Eqs.~(\ref{eq:Rx}) and (\ref{eq:threshold}) in~Subsec.~\ref{2B3} represents a particular instance of the VTEAM model, which can accurately describe experimental devices~\cite{kvatinsky2015vteam}. Within the VTEAM, the function that describes the evolution of the state (the right-hand side of Eq.~(\ref{eq:threshold})), as a function of voltage, is $\propto (V-V_{on/off})^{\alpha_{on/off}}$, where $\alpha_{on/off}$ is a constant parameter. According to~\cite{kvatinsky2015vteam}, Eq.~(\ref{eq:threshold}) is relevant to the transition from the ``on'' state to the ``off'' state in the Pt-Hf-Ti memristor~\cite{johnson2010memristive} and thus our theory can be verified on this experimental platform. Separate investigations are necessary for different $\alpha_{on/off}$ values and more complex circuits. We anticipate that when $\alpha_{on/off}>2$, the most effective approach could be the use of the maximum permissible voltage, followed by a zero voltage.

Finally, pulse control is a commonly employed technique in memristive switching. Notably, the time-dependent current/voltage waveforms introduced in this paper can be utilized as pulses, which reduces the energy needed for memristive switching compared to using pulses with a fixed amplitude. Furthermore, in digital circuit architectures, continuous voltage waveforms can be replaced by a finite set of voltage levels.

\section{Conclusion}

The production of electricity from fossil fuels is one of the main contributors to global warming. To suppress climate change, energy-efficient systems, devices, and technologies must be implemented. Our study lays the foundation for future research in optimizing the operational conditions of memristive devices, which could potentially be adapted for other memory circuit elements~\cite{diventra09a} with suitable modifications. The experimental validation of our methods on single devices~\cite{johnson2010memristive,9035621} and their application to memristive circuits, including crossbar arrays~\cite{7168603,xia2019memristive}, is of significant interest. The practical implications of this work can be significant, as implementing the protocols outlined here could enhance the efficiency of memristive systems and related technologies such as in-memory computing and neuromorphic computing, to name a few.

\section*{Acknowledgement}
The authors thank Alon Ascoli and Fernando Corinto for useful comments and discussions. YVP was supported by the NSF grant EFRI-2318139.


\end{document}